\def\ba{\begin{eqnarray}}

\def\ea{\end{eqnarray}}

\def\lb{\label}

\def\nn{\nonumber \\}

\def\bi{\bibitem}

\def\D{\Delta}

\def\ee{\epsilon}

\def\f{\phi}

\def\ff{\bar{f}}

\documentclass[12pt]{article}

\begin{document}

\title{ Hidden asymmetry and  forward-backward correlations}

\author{A.Bialas  and K.Zalewski\\ H.Niewodniczanski
Institute of Nuclear Physics\\
Polish Academy of Sciences\thanks{Address: Radzikowskiego 152, Krakow,
Poland}\\and\\
M.Smoluchowski Institute of Physics \\Jagellonian
University\thanks{Address: Reymonta 4, 30-059 Krakow, Poland;
e-mail:bialas@th.if.uj.edu.pl;}}

\maketitle

keywords: forward-backward correlations, cumulants, factorial moments

\begin{abstract}

A model-independed method of studying the forward-backward correlations
in symmetric high energy processes is developped. The method allows a
systematic study of properties of various particle sources and to
uncover  asymmetric structures hidden in symmetric hadron-hadron and
nucleus-nucleus inelastic reactions.

\end{abstract}

\vspace{0.3cm}

\section {Introduction}

New data from LHC on soft particle production in pp collisions
\cite{alice,atlas,cms} open a new chapter in the long history of this
problem thus reviving some questions which were raised already many
years ago. One of these questions concerns forward-backward correlations
which was extensively studied in the framework of specific models
\cite{kitw} and shown to be useful in discriminating between various
mechanisms of particle production.

In the present paper we develop a systematic, model-independent method to
study forward-backward correlations and show that it may be effective in
addressing the issue (debated since early seventies) of the number and
nature of quasi-independent sources of particles contributing to
particle production in various rapidity regions. Specifically,
restricting ourselves to symmetric reactions (like $pp$ or $Au-Au$
collisions) we ask what are the contributions from symmetric and
asymmetric sources and how to measure them experimentally. The problem
may be interesting since various models differ substantially in this
respect.

The simplest hypothesis is to say that there is just one symmetric
source. This is the case of the Landau hydrodynamic model \cite{landau}
and its recent modifications \cite{hydro} where particle production is
governed by evolution of a fluid. Similar conclusion follows from the
simple multiperipheral model \cite{afs} suggesting just one symmetric
source in the form of the multiperipheral chain. This idea was then
reformulated in the parton model \cite{feynman} and in the
bremsstrahlung mechanism \cite{stod}. One may of course also
consider many symmetric sources.

A more sophisticated possibility, taking into account the colour
structure of the colliding systems, was formulated in the Dual Parton
Model \cite{dpm}. Here the number of sources depends on energy of the
collision and on the type of the projectiles. For $p-p$ collisions, at
relatively low energies there are two asymmetric sources (chains),
spanned between a valence diquark from one projectile and the valence
quark from the other one. For nucleus-nucleus collisions the picture is
similar, but the number of asymmetric sources fluctuates, depending on
number of participants in the two colliding nuclei. As energy increases,
contributions from symmetric chains, spanned between the sea quarks and
antiquarks, come into play.

Simpler ideas were also put forward. In the wounded nucleon model
\cite{bbc} particles are emitted indepedently from the two colliding
nucleons thus creating two asymmetric sources \cite{bc}. A similar idea,
applied to the constituent quarks and diquarks, was proposed in
\cite{bb}. In the Fritjof model \cite{fritjof} there are also two
sources, essentially two large diffractively produced clusters (each one
related to one of the colliding hadrons).

In the present paper we consider only symmetric collisions (e.g. p-p or
Au-Au). We show that in this case a systematic study of forward-backward
correlations allows to distinguish between the various possibilities
listed above and to obtain information about some characteristic
properties of the sources.

Studies of forward-backward correlations in specific models have a long
history, see e.g. \cite{kitw,dpm}. Our work was mostly triggered by a
recent series of papers by Bzdak \cite{bzdakx}-\cite{bzdakw}, suggesting
that a strong asymmetric component is present not only in $p-p$
\cite{kitw} and $d-Au$ \cite{bc} but also in $Au-Au$ collisions. This
observation raises interesting questions about the
hydrodynamic evolution of the quark-gluon plasma believed to be produced
at RHIC \cite{bw}.

In the next section we formulate the problem in terms of generating
functions. In Section 3 the relations which allow to test various
hypotheses in $p-p$ collisions are given. Symmetric nucleus-nucleus
collisions are discussed in Section 4. Some comments  on recent STAR
measurements \cite{star1} are given in Section 5. Our conclusions are
listed in the last section.

\section {General formulation}

If particles emerge from $M$ independent sources, the generating
function for the particle distribution in a phase-space region $G$ is a
product of generating functions describing distributions of particles
from individual sources.

Restricting the discussion to rapidity spectra\footnote{Our discussion
applies to any variable symmteric with respect to an axis or a plane}, we shall
consider two sources which are asymmetric with respect to to
$y=y_{c.m.}=0$ and a third symmetric one. Their generating functions
are denoted by $\phi_L$, $\phi_R$,, $\phi_C$.

We shall discuss multiplicity distributions in
 two intervals of rapidity, denoted by $\D_L$ and $\D_R$,
situated symmetrically with respect to $y=0$.

Consider first the sum $[\D_L+\D_R]$. The generating function of the
multiplicity distribution in $[\D_L+\D_R]$ can  be written as
\ba
\Phi(z;w_L,w_R,w_C)\equiv\sum_n P(n) z^n =[\phi_L(z)]^{w_L}[\phi_R(z)]^{w_R}
[\phi_C(z)]^{w_C}.\lb{gen}
\ea
where $w_L$, $w_R$, $w_C$ are numbers of the relevant sources.

Assuming now that  the splitting
between $\D_L$ and $\D_R$ of particles  from each source is random (i.e. it
follows the binomial distribution) we have for the joint
 distribution in $\D_L$ and $\D_R$
\ba
\phi_m(z_L,z_R)\equiv\sum_{n_L,n_R}P_m(n_L,n_R)z_L^{n_L}z_R^{n_R}=
\phi_m(p_{Lm} z_L+p_{Rm}z_R)  \lb{fim}
\ea
where $p_{Lm}$ and $p_{Rm}$ are  probabilities that a particle emitted
from the source labelled  $m$ ends up in $\D_L$ or in $\D_R$. Thus
$p_{Lm}+p_{Rm}=1$. Symmetry implies
\ba
p_{LL}=p_{RR}\equiv p_+;\;\;\;p_{LR}=p_{RL}\equiv p_-=1-p_+;\;\;\;
p_{LC}=p_{RC}=\frac12. \lb{pik}
\ea
Consequently
\ba
\Phi(z_L,z_R;w_L,w_R,w_C)=\nn=
\left[\phi(p_+z_L+p_-z_R)\right]^{w_L}\left[\phi(p_-z_L+p_+z_R)\right]^{w_R}
 \left[\phi_C(z_L/2+z_R/2)\right]^{w_C}.
\lb{lrnuc}
\ea
For symmetric collisions we have
$w_L=w_R\equiv w$ and $\phi_L(z)=\phi_R(z)\equiv\phi(z)$. We
also note that the distribution of particles in one of the considered
intervals, say $\D_L$, is evaluated from $\Psi(z)=\Phi(z_L=z,z_R=1)$.

From these formulae one can evaluate all moments of the joint
distribution in $\D_L$ and $\D_R$, as well as the moments of the
distribution in one of the intervals, in terms of the moments of the
distributions describing the sources.

When only symmetric or only asymmetric sources are present, one can  derive
 relations between the joint
moments (describing the forward-backward correlations) in terms of the
moments characterizing the distribution in one of the intervals. These
relations provide demanding tests for these hypotheses,  allowing to
distinguish between various mechanisms of particle production. When both
symmetric and asymmetric sources contribute
the relations allow to obtain information on
distributions characterizing the sources themselves.

In the next two sections we discuss some of these relations.

\section {Relations between cumulants}

In this section we derive relations
between the cumulants $f_{ik}$ of the joint distributions in $\D_+$ and $\D_-$
and the cumulants $f_i$ of the distribution in one of the intervals.

In terms of the generating functions  $\Phi(z_L,z_R)$ and
$\Psi(z)$
the cumulants are defined as
\ba
f_{kl}= \frac{\partial^{k+l}\{\log \Phi(z_L,z_R)\}}{\partial^k z_L\partial^l
z_R}_{[z_L=z_R=1]};\;\;\;\;  f_{i0}\equiv
f_i=\frac{d^i \{\log\Psi(z)\}}{d z^i}_{[z=1]}. \lb{cum}
\ea
Since logarithm changes the products in $\Phi(z_L,z_R)$ and
$\Psi(z)$  into
sums, we immediately obtain
\ba
f_{k+l}-f_{kl}=\frac12[p_+^{k+l}+p_-^{k+l}-p_+^kp_-^l-p_+^lp_-^k]
\ff_{k+l}= g_kg_l\ee^2 \ff_{k+l} \lb{3s}
\ea
where
\ba
2\ff_i=\frac{d^i \{\log[\phi(z)]^{2w}\}}{d z^i}_{[z=1]}
\ea
are the cumulants of the particle distribution in $[\D_+$ +
$\D_-]$ coming from the two asymmetric sources and
\ba
\ee=p_+-p_-;\;\;\; g_k=\frac{p_+^k-p_-^k}{p_+-p_-}=\frac1{2^{k-1}}
\sum_{j=0}^{k/2}
\left(\begin{array}{c}\!\!k\\\!\!\!2j+1\!\!\! \end{array} \right)
\ee^{2j}
\ea
Note that in (\ref{3s}) the dependence on the
number of sources and contributions from the symmetric sources drop out.

The cumulants $f_i$ and $f_{kl}$ can be determined from
the standard factorial moments  $F_{i0}=F_{0i}\equiv F_i$ and from
the joint factorial moments in two intervals
\ba
F_{kl}\equiv<\!n_L...(n_L\!-\!k+\!1)n_R...(n_R\!-\!l+\!1)\!>=
 \frac{\partial^{k+l} \Phi(z_L,z_R)}{\partial^k z_L\partial^l
z_R}_{[z_L=z_R=1]}\lb{mom}.
\ea

The relevant  relations become rather involved at high orders.
The first few are listed  in the Appendix.

One sees that, generally, the result depends on two functions $\phi$,
$\phi_C$ and one parameter ($p_+$ or $p_-$). The resulting relations
involve not only the observed moments but also the moments of the
distributions characterising the asymmetric sources.
Thus, if $p_+\neq p_-$ (i.e. if the sources are indeed
asymmetric), one can - from the measured $F_{kl}$ and $F_{k+l}$ - obtain
information about $\ff_{k+l}$ characterizing the distributions of
particles from asymmetric sources.

If there are only asymmetric sources we have obviously $\ff_{k+l}=f_{k+l}$,
so that all quantities entering the relations (\ref{3s}) can be
measured. In other words, in this case (\ref{3s}) represent identities
between the measurable quantities which must be satisfied if the
symmetric sources are not present.

The  relation for $k=l=1$ is of particular interest,
 as it allows to determine
the parameter $p_+=1-p_-$ and thus to determine the distribution in
rapidity of the particles from a single asymmetric source \cite{bzdakx}.
This determination is of course valid only if the two-sources idea is
satisfied by data. The  other relations can be used to verify this
assumption

When only the symmetric source is present, all moments $\f_i$
 vanish and  we obtain $f_{kl}=f_{k+l}$, implying
\ba
F_{kl}=F_{k+l},  \lb{fsym}
\ea
a really strong constraints. Naturally, identical result is obtained
 when $p_+=p_-$, i.e. when
the asymmetric sources become symmetric.

\section {Nucleus-nucleus collisions}

Investigation of forward-backward correlations in nucleus-nucleus
collisions may be  of particular interest, as it allows to verify to
what extent the
asymmetric components survive the period of thermalization and
hydrodynamical expansion which are believed to determine the outcome of
the collision.

To apply our analysis to this case, one has to take into account that
the numer of sources may fluctuate, depending on the numer of collisions
and/or wounded nucleons \cite{bzdaky,bzdakw,frank,lmc}. Denoting the number of
left(right) movers by $w_L(w_R)$ and the number of symmetric sources by
$w_C$ we obtain for the generating function of the joint distirbution in
$\D_L$ and $\D_R$
\ba
\Phi_w(z_L,z_R)=\sum_{w_L,w_R,w_C}W(w_L,w_R,w_C)
\Phi(z_L,z_R;w_L,w_R,w_C)
\ea
where  $W(w_L,w_R,w_C)$ is  the probability distribution of the numbers
of sources and $\Phi(z_L,z_R;w_L,w_R,w_C)$ is given by (\ref{lrnuc}).
The generating function of the distribution in one of the intervals, say
$\D_L$, is $\Psi_w(z)=\Phi_w(z_L=z,z_R=1)$.

Fluctuations in number of sources imply that the generating function
$\Phi_w(z_L,z_R)$ is no longer a product of functions desribing the
sources\footnote{Needless to say that, if there are no correlations
between numbers of sources, i.e. if the probability $W(w_L,w_R,w_C)$ is a
product of three factors, the situation reduces to the one described in
the previous section.}. Therefore relations between cumulants become
rather involved. It turns out that somewhat simpler relations are
obtained for the factorial moments. For symmetric processes, where
$W(w_L,w_R,w_C)=W(w_R,w_L,w_C)$,  the simplest ones read
\ba
F_2-F_{11}=\ee^2 <\!\!L_2- L_1R_1\!\!>;\nn
F_3-F_{12}=\ee^2\left\{<\!L_3-L_2R_1\!>+<\!C_1[L_2-L_1R_1]\!>\right\};\nn
F_4\!-\!F_{22}= \ee^2\{<\!\!L_4-L_2R_2\!\!>\!
+2<\!\!C_1[L_3 -L_2R_1]\!\!>\!+\!<\!\!C_2[R_2-L_1R_1]\!\!>\};\nn
F_4-F_{13}=\ee^2\left\{(1-p_+p_-)<\!L_4\!>-\ee^2<\!L_3R_1\!\!>
-3p_+p_- <\!\!L_2R_2\!\!>\right\}+\nn+\frac32 \ee^2
<\!\!C_1[L_3-L_2R_1]\!\!>
+\frac34 \ee^2<\!\!C_2[L_2- L_1R_1]\!\!> \lb{facm}
\ea
where $<...>$ denotes the average over the number of sources and
\ba
L_i=\frac{d^i\left\{[\phi(z)]^{w_L}\right\}}{dz^i}_{[\!z\!=\!1\!]};\;\;
R_i=\frac{d^i\left\{[\phi(z)]^{w_R}\right\}}{dz^i}_{[\!z\!=\!1\!]};\;\;
C_i=\frac{d^i\left\{[\phi_C(z)]^{w_C}\right\}}{dz^i}_{[\!z\!=\!1\!]}. \lb{ws}
\ea
are moments of the distribution in $\D_L+\D_R$ produced by left, right
and symmetric sources, respectively. They cannot be directly measured,
so (\ref{facm}) can be used to obtain information about them. Such
analysis faces, however, a difficulty: One sees that the R.H.S. of
(\ref{facm}) depends not only on the structure of the sources but also on
correlation between the numbers of various sources. To disentagle these
two effects it is necessary to study processes with various nuclei
and at  various centralities \cite{bzdaky}.

If the symmetric  sources are absent, (\ref{facm}) simplifies to
\ba
F_2-F_{11}=\ee^2[ <L_2>- <L_1R_1>];\nn
F_3-F_{12}=\ee^2[<L_3>-<L_2R_1>];\nn
F_4-F_{22}= \ee^2[<L_4>-<L_2R_2>];\nn
F_4-F_{13}=\ee^2[(1-p_+p_-)<L_4>-\ee^2<L_3R_1>
-3p_+p_- <L_2R_2>]. \lb{facmas}
\ea

Finally, if only symmetric sources are present we return to the simple
relation (\ref{fsym}).

\section {Three intervals}

Fluctuations of the number of sources in nucleus-nucleus collisions are
difficult to control because even precise determination of centrality of
the collision is not suffcient to guarantee a fixed number of sources
\cite{bzdaky,bzdakw,frank,lmc}. To improve this,  STAR collaboration
measured correlations in three intervals \cite{star1}. Apart from
the $\D_L$ and $\D_R$, one adds the third interval $\D_C$, located
centrally around $y_{cm}=0$ and not overlapping with $\D_L$ and $\D_R$.
The correlations between $\D_L$ and $\D_R$ are measured under the
constraint that a fixed number of particles, $n_C$, was found in $\D_C$.
Particle multiplicity in $\D_C$ is obviously related to the number of
sources and therefore one may expect it to be helpful in estimating this
number on event-by-event basis.

The extensive general discussion of this data was given in \cite{lmc}.
They were also analyzed in \cite{bzdaky} and \cite{frank} within
specific models.

We would like to add three comments.

(i) The relations derived in previous sections remain intact if one adds
the condition that a certain number of particles is observed in the
central interval $\D_C$. This should be clear from the derivation:
replacing the probabilities in (\ref{gen}) and (\ref{fim}) by
conditional probabilities (fixing the number of particles in $\D_C$)
does not change at all our argument.

(ii) When  symmetric and asymmetric sources are present, restricting
$n_C$ has only a limited effect on reduction of fluctuations of the number
of sources. This can be seen by considering the distribution of
particles in $\D_C$. The relevant generating function is
\ba
\Phi_C(z_C)=\left[\phi(1-p_C+p_Cz_C)\right]^{w_L+w_R}
 \left[\phi_C(1-q_C+q_Cz_C)\right]^{w_C}. \lb{fic3}
\ea
where $p_C$ is the probability that a particle from an asymmetric source
lands in $\D_C$, and $q_C$ is the probability that a particle from
a symmetric source lands there.

One sees from (\ref{fic3}) that the particle distribution in $\D_C$,
does not depend on the difference $w_-=w_L-w_R$ and thus fluctuations of
$w_-$ are not restricted by (\ref{fic3}). Moreover, (\ref{fic3}) implies
that the distribution of $n_C$ depends on both $w_+=w_L+w_R$ and $w_C$.
Since the forward-backward correlations induced by fluctuations of
asymmetric sources are, generally, weaker than those induced by the
symmetric ones, its is important to separate the two contributions. We
conclude that although measurements at a fixed $n_C$ may be helpful,
probably some additional model assumptions are necessary to disentangle
this problem.

Let us also note that in \cite {frank,lmc} only the symmetric sources
were discussed and therefore this aspect of the problem did not appear.

(iii) Using the methods of sections 3 and 4, relations can also be
derived for the joint moments of the distribution in the three
intervals. The only difference is that one has to consider the
generating function of three variables.

\section {Summary and comments}

A systematic, model-indepedent method of studying the forward-backward
correlations in particle production is developped. It is shown that it
provides a useful tool for determining the structure of the sources of
particles created in high energy cllisions. In particular, it may be
used to uncover left-right asymmetric components present in these
processes and to study their properties. This point is of interest
since, as explained in Introduction, various mechanism of particle
production differ in their predictions for the presence and/or intensity
of such asymmetric sources both in ($p-p$) and {$A-A$} collisions.

The existing data \cite{ua5} show reasonable agreement with the idea
that just two asymmetric sources dominate the observed correlations in
$p-p$ collisions \cite{kitw,dpm,bzdak}. A similar conclusion was
obtained recently in the analysis of the $Au-Au$ collisions \cite
{phob}, where data could also be explained without any symmetric
contribution being present \cite{bzdakw}. This seems not to be the case
\cite{lmc} for the more restrictive STAR data \cite
{star1}\footnote{See, however, \cite{bzdaky,frank}.}. These
conclusions have of course important consequences for selecting the
possible mechanisms of particle production.

As was indicated in Section 5, the analysis of the heavy ion data may
require additional information about the correlation between the various
particle sources. Even in this complicated situation the method proposed
here can, however, clearly distinguish whether symmetric or asymmetric
sources dominate the process in question.

New data from LHC \cite{alice,atlas,cms} show that the multiplicity in
the central rapidity region increases with energy much faster than
expected from simple extrapolations of the trends observed at lower
energies. One possible explanation is that, as predicted in some models
\cite{dpm}, a new symmetric source of particles may be excited at these high
energies. Studying the forward-backward correlations using methods
developped in the present paper should be helpful in verification of
this idea and, possibly, identification of this new component, as well
as in investigation of its properties.

A question which may be studied by the methods proposed in this paper,
is the comparison of the forward-backward correlations in $p-p$ and in
heavy ion collisions. Such comparative studies at LHC energies would
allow to obtain information about {\it longitudinal} dynamics of
quark-gluon plasma, the problem which is barely touched by the existing
analyses. Forward-backward correlations are created at the very early
stage of the collision (see, e.g. \cite{lmc}) and, apparently, survive
the evolution of the system. It remains, however, an interesting
question to what extent they are distored during this evolution.

Finally, let us emphasize that our approach is not restricted to
rapidity distributions: it can be used to study correlations in other
variables as well. One interesting possibility is to repeat the standard
analysis in rapidity restricting, however, the azimuthal angle (keeping
of course forward-backward symmetry). This may provide useful
 information on correlations in the directed flow.

Another attractive possibility is to consider correlations in tranverse
momentum at a fixed and/or opposite rapidity. Since the $p_t$
distribution in minimum bias sample is now becoming accessible in a
rather broad range \cite{alice,atlas,cms}, the full potential of the
method can be explored. Such investigation may help in disentangling the
jet structure in the low and medium $p_t$ regions.

\section{Appendix}

Some explicit  relations between  cumulats and factorial moments
(for symmetric collisions) are listed below:
\ba
f_1=F_1;\;\;\; f_{11}=F_{11}-F_1^2;\;\;\;
f_{21}=F_{21}-2F_{11}F_1-F_2F_1+2F_1^3;\nn
f_{22}=F_{22}-4F_{21}F_1-2F_{11}^2-F_2^2+4\left[F_2+2F_{11}\right]F_1^2-6F_1^4;\nn
f_{31}=F_{31}-3F_{21}f_1-F_3F_1-3F_{11}F_2+6[F_{11}+F_2]F_1^2-6F_1^4;\nn
f_2=F_2-F_1^2;\;\;\;f_3=F_3-3F_2F_1+2F_1^3;\;\;\;\nn
f_4=F_4-4F_3F_1-3F_2^2+12 F_2F_1^2-6F_1^4.
\ea

{\bf Acknowledgements} We are thankful to Adam Bzdak for numerous
discussions which were at the origin of our interest in this problem and
for a critical reading of the manuscript. Comments by Krzysiek
Fialkowski and Wojtek Florkowski are highly appreciated. This
investigation was supported in part by the grant N N202 125437 of the
Polish Ministry of Science and Higher Education (2009-2012).

\end{document}